\documentclass[sigconf]{acmart}
\AtBeginDocument{%
  }

\setcopyright{acmlicensed}
\copyrightyear{2025} 
\acmYear{2025}
\setcopyright{acmlicensed}\acmConference[KDD '25]{Proceedings of the 31st ACM
SIGKDD Conference on Knowledge Discovery and Data Mining V.2}{August 3--7,
2025}{Toronto, ON, Canada}
\acmBooktitle{Proceedings of the 31st ACM SIGKDD Conference on Knowledge
Discovery and Data Mining V.2 (KDD '25), August 3--7, 2025, Toronto, ON, Canada}
\acmDOI{10.1145/3711896.3737029}
\acmISBN{979-8-4007-1454-2/2025/08}

\newcommand{\ie}{\emph{i.e., }}

\newcommand{\hyz}[1]{\iffalse #1 \fi}
\newcommand{\modified}[1]{{\color{black}{#1}}} %

\newcommand{\za}[1]{{\color{black}{#1}}} %
\newcommand{\ours}{LLM2Rec\xspace}

\usepackage{appendix}
\usepackage{enumitem}
\usepackage{multirow}
\usepackage{makecell}
\usepackage{amsthm}
\usepackage{amsfonts}
\usepackage{url}
\usepackage{lipsum}
\usepackage{commath}
\usepackage{graphicx}
\usepackage{booktabs}
\usepackage{xspace}
\usepackage{tikz}
\usepackage{graphicx}
\usepackage{algorithm}
\usepackage{algorithmicx}
\usepackage{algpseudocode}
\usepackage{amsmath}

\begin{document}

\title{LLM2Rec: Large Language Models Are Powerful Embedding Models for Sequential Recommendation}

\author{Yingzhi He}
\email{heyingzhi@u.nus.edu}
\affiliation{%
  \institution{National University of Singapore}
  \city{Singapore}
  \country{Singapore}
}
\authornote{Equal Contribution.}

\author{Xiaohao Liu}
\email{xiaohao.liu@u.nus.edu}
\affiliation{%
  \institution{National University of Singapore}
  \city{Singapore}
  \country{Singapore}
}
\authornotemark[1]

\author{An Zhang}
\email{an.zhang3.14@gmail.com}
\affiliation{%
  \institution{University of Science and Technology of China}
  \city{Hefei}
  \country{China}
}
\authornote{Corresponding Author.}

\author{Yunshan Ma}
\email{ysma@smu.edu.sg}
\affiliation{%
  \institution{Singapore Management University}
  \city{Singapore}
  \country{Singapore}
}

\author{Tat-Seng Chua}
\email{dcscts@nus.edu.sg}
\affiliation{%
  \institution{National University of Singapore}
  \city{Singapore}
  \country{Singapore}
}


\begin{abstract}
Sequential recommendation aims to predict users’ future interactions by modeling collaborative filtering (CF) signals from historical behaviors of similar users or items.
Traditional sequential recommenders predominantly rely on ID-based embeddings, 
which capture CF signals through high-order co-occurrence patterns.
However, these embeddings depend solely on past interactions, lacking transferable knowledge to generalize to unseen domains.
Recent advances in large language models (LLMs) have motivated text-based recommendation approaches that derive item representations from textual descriptions.
While these methods enhance generalization, they fail to encode CF signals—\ie latent item correlations and preference patterns—crucial for effective recommendation. 
We argue that an ideal embedding model should seamlessly integrate CF signals with rich semantic representations to improve both in-domain and out-of-domain recommendation performance.

To this end, we propose \textbf{LLM2Rec}, a novel embedding model tailored for sequential recommendation, integrating the rich semantic understanding of LLMs with CF awareness. 
Our approach follows a two-stage training framework: (1) Collaborative Supervised Fine-tuning, which adapts LLMs to infer item relationships based on historical interactions, and (2) Item-level Embedding Modeling, which refines these specialized LLMs into structured item embedding models that encode both semantic and collaborative information.
Extensive experiments on real-world datasets demonstrate that LLM2Rec effectively improves recommendation quality across both in-domain and out-of-domain settings. 
Our findings highlight the potential of leveraging LLMs to build more robust, generalizable embedding models for sequential recommendation. 
Our codes are available at \url{https://github.com/HappyPointer/LLM2Rec}.
\end{abstract}




\maketitle

\section{Introduction}

Sequential recommendation \za{aims to predict users' future interactions by learning high-quality item representations that effectively capture both user preference patterns and item inherent content \cite{MoRec_new, AlphaRec}.}
\za{Conventional} sequential recommenders \za{typically} assign unique \za{identifiers} (IDs) to items and learn corresponding representations \za{based on historical user} interaction sequences \cite{SASRec, GRU4Rec, BERT4Rec, Caser, S3-rec, CL4SRec}. 
\za{These ID-based representations primarily encode collaborative filtering (CF) signals by solely modeling multi-hop co-occurring patterns in sequential trajectories \cite{NCF, NGCF}.
While effective, such recommenders lack item content information, making them highly domain-dependent and unable to generalize to unseen items or new domains \cite{cross_domain_challenges, yuan2020parameter, MoRec_new}.
We argue that high-quality item representations in sequential recommender systems must simultaneously encapsulate item semantics and CF signals.}

\za{
Recent advances in large language models (LLMs) have motivated extensive research into leveraging rich semantic information for improved item representation learning, including purely text-based representations \cite{AlphaRec, UnisRec, RecFormer} and hybrid representations that fusing semantic and CF signals \cite{RLMRec, LLM-ESR, BLAIR}.
Purely text-based recommenders extract item representations from pre-trained language models, offering strong generalization capabilities but disregarding CF signals.
To mitigate this limitation, hybrid methods attempt to integrate both semantic and CF signals through various fusion strategies, including directly concatenating ID-based and semantic representations \cite{MoRec_new, MMGCN, SLMRec}, guiding ID-based representation learning with content features \cite{RLMRec, CLCRec}, adopting hybrid fusion architectures \cite{LLM-ESR, LLMEmb}, and tuning embedding models to bridge CF and semantic spaces \cite{EasyRec, BLAIR}.
However, fusion-based techniques, whether simple concatenation or sophisticated mechanisms like cross-attention, struggle to learn a unified representation space, leading to misalignment between item semantics and user behavior spaces.
We believe that developing a general embedding model for sequential recommendation-one that inherently captures both CF and semantic knowledge-is a more promising yet underexplored research direction.
\modified{While recent efforts have attempted to construct unified embedding models that align item semantics with user behavior spaces through contrastive pre-training~\cite{BLAIR, EasyRec},} these methods typically require large-scale training samples and prohibitively large batch sizes to effectively encode CF signals.
More critically, they fail to fully leverage the strong semantic understanding and reasoning capabilities of state-of-the-art LLMs, such as Qwen \cite{Qwen2} and Llama \cite{llama}.
This motivates us to investigate how LLMs can be adapted to serve as generalizable embedding models for recommendation.
}

\begin{figure}[t]
    \centering
    \includegraphics[width=0.49\textwidth]{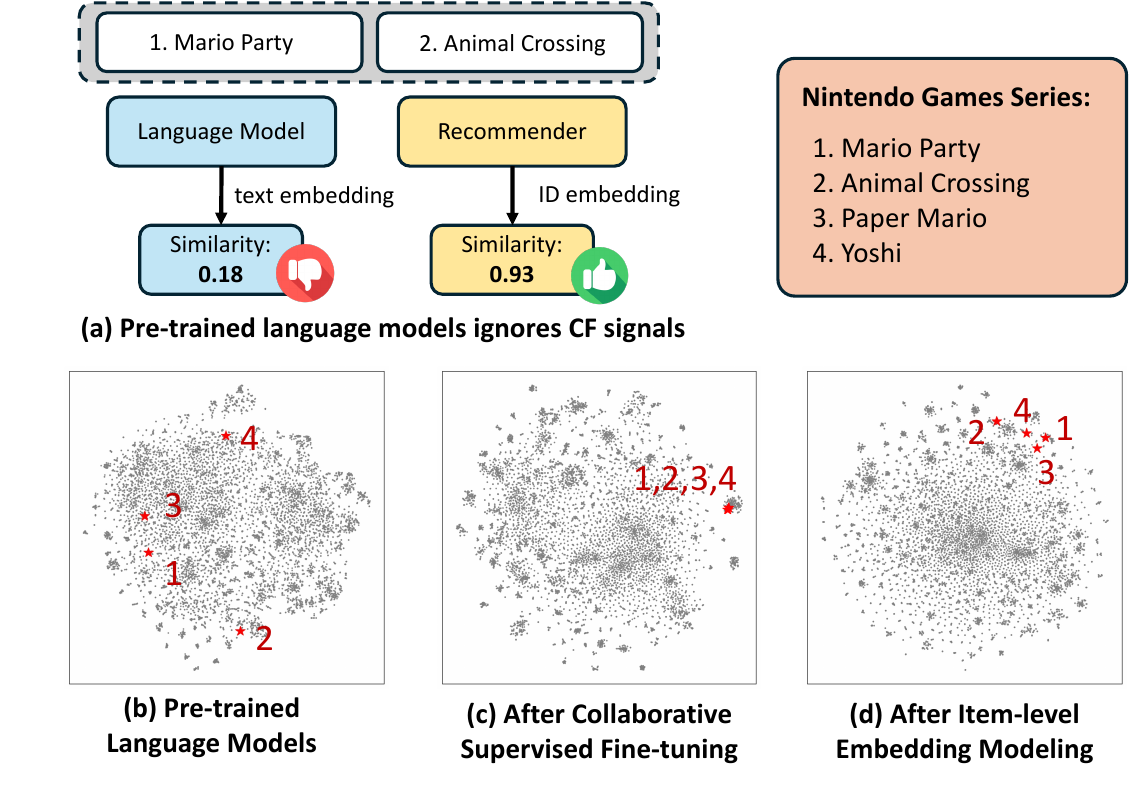}
    \vspace{-0.2in}
    \caption{An illustrating example that pre-trained language models ignore CF signals and the embedding distributions in different training stages of \ours. (a) The behaviorally correlated Mario Party and Animal Crossing are close in ID embeddings but distinct in text embeddings. (b) Pre-trained language models failed to discover the behaviorally similar items. (c) Collaborative Supervised Fine-Tuning enables the LLM to capture the CF signals. (d) Item-level embedding modeling preserves CF signals while producing more distinguishable embeddings.}
    \vspace{-0.2in}
    \label{fig:motivation}
\end{figure}

To develop a recommendation-specialized embedding model with strong generalization to unseen domains, we aim to integrate the powerful semantic understanding capability of LLMs with the ability to capture CF signals.  
Motivated by recent studies demonstrating that LLMs can effectively learn recommendation tasks through supervised fine-tuning \cite{TallRec, BigRec, InstructRec, Rella, liu2024once, SDPO}, we leverage this approach to make LLMs aware of the CF signals from user interaction sequences. 
To further facilitate the transition of the LLM from token-level prediction to item-level embedding generation \cite{LLM2Vec, NV-Embed}, we refine the CF-aware LLM into a structured recommendation embedding model with additional item-level embedding modeling.

To this end, we introduce \ours, a recommendation embedding model built upon the LLM that is explicitly aware of the CF signals. 
Specifically, our training framework includes two stages: (1) Collaborative Supervised Fine-Tuning (CSFT) and (2) Item-level Embedding Modeling (IEM). In the first stage, CSFT fine-tunes the LLM on a mixture of six real-world recommendation datasets, enforcing it to predict the next item based on the historical interaction sequence. As illustrated in Figure~\ref{fig:motivation}, the embeddings of several games in the Nintendo series are initially scattered. As these games are frequently co-purchased by users with similar preferences, their embeddings become closely clustered after CSFT. This shift indicates that the LLM learns to capture CF signals through CSFT.
In the second stage, we enable bidirectional attention with Masked Next Token Prediction (MNTP) and apply item-level contrastive learning to further facilitate the LLM to be an embedding model. Bidirectional attention enables capturing contextualized information within item titles \cite{LLM2Vec} and MNTP helps the LLM adapt to the newly introduced bidirectional attention mask. Item-level contrastive learning explicitly shifts the pre-training objective from token-level to item-level, helping to generate distinguishable item embeddings and yet preserve the CF signals. Both MNTP and item-level contrastive learning are lightweight adaptations incurring slight computational costs while remaining effective. As presented in Figure~\ref{fig:motivation}, the embeddings of Nintendo games remain close while becoming more differentiated, offering richer and more effective information for recommendation.

To assess the effectiveness of \ours, we conduct extensive experiments on both in-domain and out-of-domain datasets using various downstream sequential recommenders. Experimental results show that \ours consistently outperforms the existing embedding models across both in-domain and out-of-domain datasets. Additionally, the generalization ability of the embedding model benefits from training on mixed datasets spanning diverse categories. These findings underscore the potential of LLMs as inherently powerful embedding models for sequential recommendation.

\section{Related Work}
In this section, we briefly review the works related to this paper from two main categories: 1) sequential recommendation, and 2) embedding models.

\subsection{Sequential Recommendation}
Sequential recommendation learns item representations to predict items that users are likely to interact with in the future. From the item representation learning perspective, existing methods can be broadly categorized into three paradigms: ID-based methods, pure text-based methods, and hybrid ID-text methods.

\noindent\textbf{ID-based methods} assign each item a unique identifier and learn the corresponding representation using various sequence modeling techniques, like recurrent neural networks \cite{GRU4Rec}, convolutional neural networks \cite{Caser}, and transformer-based architectures \cite{SASRec, S3-rec}. With these techniques, ID-based methods capture item correlations and user interests from user interaction sequences. Despite the effectiveness, these methods are not capable of handling tasks from unseen domains or unseen items, lacking generalizability. 

\noindent\textbf{Pure text-based methods} represent items with text embeddings derived from item contents, such as titles or profiles, using pre-trained language models as text encoders. Within the unified language space, these methods have the potential to generalize to unseen domains~\cite{AlphaRec, UnisRec, RecFormer}.
However, their effectiveness is largely limited by the adopted embedding model due to their heavy reliance on text embeddings. Moreover, these pre-trained language models are general for language tasks rather than specialized for recommendation, resulting in suboptimal performance as well.

\noindent\textbf{Hybrid ID-text methods} generate representation by incorporating both ID and textual information. Common techniques include (1) using text embeddings to guide or enhance the learning of ID representations \cite{RLMRec, CLCRec, LLMEmb, KAR, liu2024practice, LLMRec}, (2) concatenating text and ID embeddings \cite{MoRec_new, MMGCN, SLMRec, Elimrec}, and (3) fusing ID and text information through attention architectures \cite{LLM-ESR}. While these methods leverage textual information to improve representation learning, they still rely on ID-based embeddings, indicating that their effectiveness remains highly dependent on dataset-specific training. As a result, similar to purely ID-based methods, hybrid approaches must be trained on the target domain to learn effective representations, limiting their ability to generalize to new domains \cite{cross_domain_challenges, yuan2020parameter, MoRec_new}.



\subsection{Embedding Models}

Pre-trained embedding models play a fundamental role in various downstream tasks, including information retrieval \cite{retrieval_1, retrieval_2, retrieval_3}, text similarity \cite{similarity_1}, and classification \cite{classification_1_imagenet, classification_2}. Existing approaches can be broadly categorized into two main types: 1) Language encoder models with bidirectional attention and 2) LLM-based embedding models. 
Language encoder models have long been the dominant approach for learning text embeddings. These models leverage transformer architectures with bidirectional attention, allowing them to capture richer contextual relationships and produce more effective sentence embeddings \cite{BERT, Roberta, Bart}. 
Their training objectives include next sentence prediction \cite{BERT}, masked language modeling \cite{BERT, Roberta}, and contrastive learning techniques \cite{Simcse, wang2022text, GTR, li2023towards}, which has gained significant popularity for its effectiveness in producing high-quality sentence embeddings.
LLM-based embedding methods have gained increasing popularity with the growing capabilities of large language models (LLMs).
The straightforward approaches generate sentence embeddings directly from the last hidden states of decoder-only LLMs, either by using the hidden state of the EOS token or by average pooling the hidden states of all tokens in the sentence \cite{muennighoff2022sgpt, wang2023improving, springer2024repetition}. However, directly using pre-trained LLMs as embedding models without additional training leads to suboptimal performance, as they are optimized for predicting future tokens rather than generating holistic sentence representations. To address this limitation, recent approaches focus on adapting LLMs into dedicated embedding models with further adaption \cite{LLM2Vec, GRIT, BGE-li2024, NV-Embed, Gecko, GTE}. Common techniques include sentence-level contrastive learning and attention mask modifications, both of which enhance LLMs' effectiveness as embedding models.

In recommendation tasks, the implicit relationships between items, known as CF signals, are also important alongside semantic understanding. An effective embedding model for recommendation should integrate both to maximize performance.
While most sequential recommenders rely on general embedding models, some initial efforts attempt to develop recommendation-specific embeddings. Blair~\cite{BLAIR} aligns representations of user reviews with item titles using contrastive learning on over 30 million instances across 33 categories from the Amazon platform.
EasyRec \cite{EasyRec} aligns user representations with their interacted items and further incorporates diverse user and item profiling in contrastive learning.
However, both methods constrain their backbone embedding model to smaller ones like BERT~\cite{BERT} or RoBERTa \cite{Roberta} due to the high computational cost of contrastive learning, which requires large batch sizes and sufficient training iterations.
\modified{Recently, LLMEmb~\cite{LLMEmb} extends recommendation-specific embedding models by leveraging large language models. It adopts attribute-level augmentations and aligns augmented views of the same item to enhance the generated embeddings. LLMEmb primarily treats LLMs as powerful semantic encoders and does not explicitly integrate CF signals into the learned embeddings.}
In contrast, our approach fuses CF signals into LLMs through collaborative supervised fine-tuning. With the following enhancements of lightweight item-level embedding adaptation, \ours achieves superior recommendation-specific embeddings while maintaining computational efficiency.

\begin{figure*}[t]
    \centering
    \includegraphics[width=0.98\textwidth]{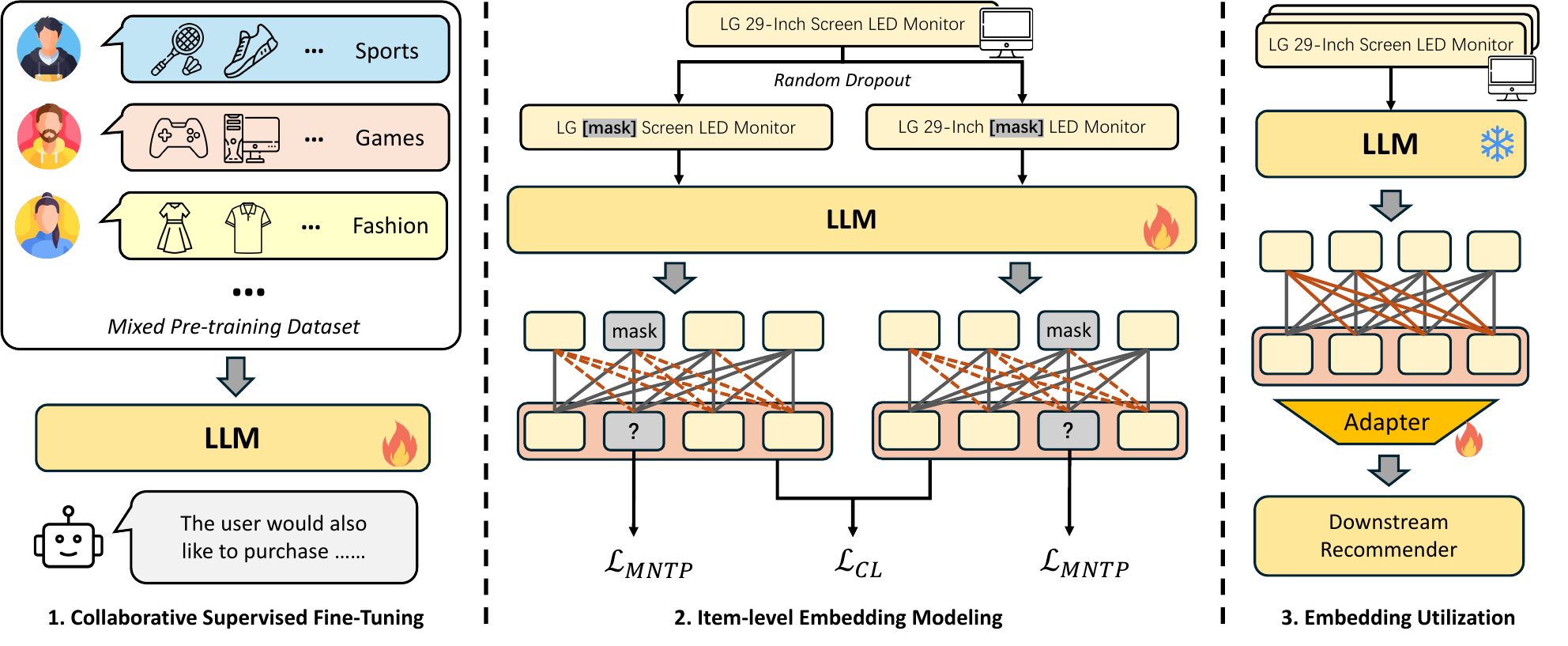}
    \vspace{-0.1in}
    \caption{Illustration of the overall pre-training framework of \ours and how the generated embeddings are utilized for downstream sequential recommenders. \ours employs a two-stage training strategy: first, adapting LLMs to infer item relationships from previous interactions, namely, collaborative instruction fine-tuning (left); second, reforming specialized LLMs for item-level embeddings with two training objectives (middle). 
    By encoding both semantic and CF information, generated embeddings bolster the exiting recommenders via a lightweight adapter (right). 
    }
    \vspace{-0.1in}
    \label{fig:framework}
\end{figure*}

\section{Methodology}

In this section, we first elaborate on the problem formulation of utilizing embedding models for sequential recommendation tasks. After that, we introduce the crux of LLM2Rec, consisting of collaborative supervised fine-tuning and item-level embedding modeling. At the end, we elaborate the optimization of LLM2Rec, followed by its utilization for downstream sequential recommenders. The overall framework is depicted in Figure~\ref{fig:framework}.

\subsection{Problem Formulation}

\subsubsection{Sequential Recommendation}
We have a set of items $\mathcal{I}$, and user interaction sequence $X$, constructing the dataset $\mathcal{D} = \{\mathcal{I}, X\}$.
Wherein, each element denotes an item sequence $\{i_1, i_2, \dots, i_{N_u}\}$; $N_u$ is the number of the interaction items for a specific user $u$. 
In our setting, these items are typically represented by their titles.  
The goal of sequential recommendation is to predict the next item $i_{N_u + 1}$ given the previous interactions $i_{<N_u}$. 
Here we hope to obtain a sequential recommender $R(i_{<N_u})$ that is capable to capture the user intention hidden within the previous interaction and finally inferring the preferred next item. Formally, we denote it as
$
p(i_{N_u + 1} | i_{<N_u}) = R(i_{<N_u}).
$

\subsubsection{Recommendation Embedding Modeling}
In this work, we focus on elevating the sequential recommenders via adapting a well-trained embedding model $\mathcal{E}(\cdot)$.
Given training datasets $\mathcal{D}^{\text{train}} = \{\mathcal{D}_k\}_{k=0}^{n}$, the embedding model is capable of generating effective embedding for testing datasets (\textit{i.e.}, out-of-domain datasets) $\mathcal{D}^{\text{test}} = \{\mathcal{D}_{n+j}\}_{j=0}^{m}$. 
$\mathcal{E}(\cdot)$ can accept the textual description of item sequence, \textit{i.e.}, $\{\mathbf{{t}}_i:=[{t}^i_1, {t}^i_2, \dots, {t}^i_{\ell_i}]\}_{<N_u}$), or a single item; $\ell_i$ the token length of item $i$. 
Accordingly, leveraging such an embedding model, we can obtain item embedding via $\mathbf{z}_i = \mathcal{E}(\mathbf{{t}}_i)$. 
Note that, in text-based recommendation, the above methods typically adopt a pre-trained text encoder $\mathcal{E}^*(\cdot)$ to extract the semantics in a latent space. 
Thus, we can reformulate the sequential recommendation by explicitly involving the embedding modeling:
\begin{align}
    p(i_{N_u + 1} | i_{<N_u}) = R({\mathbf{z}_i}_{<N_u}), \forall\mathbf{z}_i = \mathcal{E}(\mathbf{i}).
\end{align}

\subsection{Collaborative Supervised Fine-tuning}
We adapt LLM for sequential recommendation tasks via supervised fine-tuning with collaborative information, \textit{i.e.,} user-item interactions.  First, we construct the recommendation instructions that take the user previous interactions as input, and set the next item as the label. 
This formulation follows the LLM's inherent autoregressive generation manner; and our goal is to enable LLMs to perform recommendations with collaborative instructions.  

\begin{figure}[h]
    \centering
    \begin{tikzpicture}
        \node[draw, fill=gray!10, text width=0.9\linewidth, align=left, inner sep=0.04\linewidth] at (0, 0) { 
            {\textbf{Input:} \\
            \hspace{1.5em} Logitech G13 Gameboard, \\
            \hspace{1.5em} \colorbox{red!10}{\textbf{Tamron 70-200mm Camera Lens (Nikon)}}, \\
            \hspace{1.5em} Pelican SD Card Case, \\
            \hspace{1.5em} \colorbox{red!18}{\textbf{YONGNUO Flash Trigger (Canon)}}, \\
            \hspace{1.5em} TAKSTAR SGC-598 Microphone, \\
            \hspace{1.5em} \colorbox{red!10}{\textbf{STK EN-EL3e Charger for Nikon Camera}}, \\
            \hspace{1.5em} VGA to HDMI Cable, \\
            \hspace{1.5em} Allstate 2-Year Protection Plan, \\
            \hspace{1.5em} BLACKRAPID Lock Star Cover.
        } \\
            
            \vspace{0.25cm}
            \hrule height 0.5pt
            \vspace{0.25cm}
            \textbf{Output:} {Neewer Wireless Flash Trigger for Camera}
        };
    \end{tikzpicture}
    \vspace{-0.1in}
    \caption{An example of the collaborative instruction for fine-tuning. \iffalse Item titles are simplified for better readability.\fi}
    \label{fig:prompt}
    \vspace{-0.2in}
\end{figure}

As aforementioned, we represent items with textual information (\textit{e.g.}, titles), denoted as $\mathbf{{t}}_i:=[{t}^i_1, {t}^i_2, \dots, {t}^i_{\ell_i}]$, where $t$ is the token indices; the user interactions can be represented as the concatenated item sequence, $\mathbf{{t}}_u := [\mathbf{i}_1, \mathbf{i}_2, \dots, \mathbf{i}_{N_u}]$. 
As shown in Figure~\ref{fig:prompt}, the input instruction is a series of items (\textit{i.e.}, movie titles), followed by the next item title as the desired output. The CF signals manifest by these highlighted item correlations (the darker the color, the more similar with the target items). To mitigate the influence of the template, which introduces hidden states irrelevant to the current item representation, we only retrain the item titles with some necessary separators, like commas. 
With such type of collaborative instructions,  we employ the objective to predict the tokens within the next item autoregressively:
\begin{align}
    \mathcal{L}_{\text{CSFT}} = -
    \sum_{(u,i)\in X} \sum^{\ell_i}_{s=0} p(\mathbf{{t}}_{i,s} | \mathbf{{t}}_u ,\mathbf{{t}}_{i, <s} ),
    \label{eq:csft}
\end{align}
where $s$ denotes the current step. 
This step is simple yet essential to unlock the capability of LLM for recommendation and bolsters the following embedding modeling.

\subsection{Item-level Embedding Modeling}

Decoder-only LLMs are designed for autoregressive prediction, making them less effective at generating high-quality embeddings, a task typically suited for encoder models.
However, LLMs exhibit stronger semantic understanding ability compared to traditional bidirectional encoder methods~\cite{NV-Embed, LLM2Vec}.
Some works~\cite{zhang2024notellm, zhang2024laser, chen2024hllm} propose utilizing the last hidden state of the token to represent the given text, serving as an alternative to use LLMs.
However, we hope to reap both benefits of decoder-only LLM's knowledge and encoder models' architecture, thus reforming LLMs for item-level embedding modeling.

\subsubsection{Reforming Decoder-only LLM to Encoder}
The key differences that distinguish the encoder and decoder language models lie in \textit{causal attention} and \textit{training objective}.  Inspired by remarkable language encoders~\cite{BERT, BGE-li2024} and recent attempts on utilizing LLM as general embedding models~\cite{GTE, LLM2Vec}, we equip the fine-tuned LLMs with \textit{bidirectional attention} and \textit{masked next token prediction optimization} (MNTP).

\noindent\textbf{Causal $\to$ Bidirectional attention.}
Causal attention restricts access to information from later tokens when generating token embeddings. While this is essential for prediction tasks, embedding models require visibility of both preceding and succeeding tokens to capture comprehensive contextual details.
Motivated by this, we cancel the causal attention mask, enabling prediction based on the item-level past and future context (\textit{i.e.}, bidirectionally). 
To adapt model parameters for the newly introduced architecture, we impose an additional training stage with masked next token prediction (MNTP). 
Given an item sequence as input, we randomly mask tokens with a pre-defined fraction, then train LLMs with masked next token prediction task:
\begin{align}
    \mathcal{L}_{\text{MNTP}} = -
    \sum_{i\in \mathcal{I}} \sum^{\ell_i}_{s=0} p(\mathbf{{t}}_{i,s} | \mathbf{{t}}_{i, <s}).
    \label{eq:mntp}
\end{align}
The parameters of LLM that are trained with sequential recommendation tasks in causal attention, followed by MNTP tasks in bidirectional attention. 
Notably, this stage focuses solely on the information within a single item, aligning with the tuning of LLMs for item-level embeddings. In contrast, CSFT captures relationships between different items within user interaction sequences.

\subsubsection{Item-level Contrastive Learning}
Token-level embedding modeling is developed with architecture modification and MNTP optimization; however, we hope to generate item-level embedding, which is more prevailing and intuitive for downstream recommenders. The straightforward solution can be a direct average-pooling of token-level embeddings. 
Specifically, for item $i$, the item embeddings can be represented as $\mathbf{z}_i = 1/\ell_i \sum_{j\in[\ell]} \mathbf{z}_i^{j}$.
Hence, the embedding model can be the composition of such average pooling operation and our modified LLM model $\pi_\theta$,  and formally denoted as $\mathcal{E} := \text{avg}\circ \pi_\theta$. 
In this work, we further enhance this by employing item-level contrastive learning. 

\noindent\textbf{Token $\to$ Item level embedding. }
The input item $\mathbf{t}_i$ is passed through LLM model twice with random masking independently, yielding two views of the same item (\textit{i.e.}, $\mathbf{\mathbf{\tilde{t}}_i^1}$ and $\mathbf{\mathbf{\tilde{t}}_i^2}$). 
Following the unsupervised contrastive learning paradigm~\cite{Simcse}, we optimize the parameters via: 
\begin{align}
    \mathcal{L}_{\text{IC}} = -
    \sum_{i\in \mathcal{I}} 
    \log \frac{\left( \mathcal{E}(\mathbf{\mathbf{\tilde{t}}_i^1})\mathcal{E}(\mathbf{\mathbf{\tilde{t}}_i^2})^\top /\tau\right)}{\sum_{j\in\mathcal{I}} \left(\mathcal{E}(\mathbf{\mathbf{\tilde{t}}_j^1})\mathcal{E}(\mathbf{\mathbf{\tilde{t}}_j^2})^\top /\tau\right)}, 
    \label{eq:ic}
\end{align}
where $\tau$ is the temperature ratio.
With this objective, item embeddings are learned via a holistic view and contrasted with others to enhance their distinctiveness. The differentiation induced by contrastive learning aligns well with a series of recommendation methods~\cite{wu2021self, CL4SRec}, providing a stronger foundation for downstream sequential recommenders.

\subsection{Optimization \& Utilization}

\noindent\textbf{Training LLM2Rec.} We train the LLMs in a sequential manner, from collaborative supervised fine-tuning ($\mathcal{L}_\text{CSFT}$) to item-level embedding modeling ($\mathcal{L}_\text{MNTP}$ and $\mathcal{L}_\text{IC}$). The model architecture is modified in the second stage, reforming causal attention to the bidirectional. 
This training strategy progressively enhances the LLMs' capabilities, featuring them to capture both semantic and CF information for recommendation.

\noindent\textbf{Empowering downstream recommenders.}
After the training of LLM2Rec, we introduce the utilization of embeddings generated from LLM2Rec to bolster the existing sequential recommenders. 
Most sequential recommenders obtain their item embeddings from scratch or initialize them as trainable parameters updated along with the training~\cite{SASRec, GRU4Rec}. 
In this work, we provide a simple solution via an additional linear adapter: $\mathbf{z}'_i = \mathbf{w}\mathbf{z}_i +\mathbf{b}$, where $\mathbf{w}$ and $\mathbf{b}$ are the weight and bias matrices. 
We use these transformed embeddings as item representations.
Notably, the parameters is optimized through downstream recommenders' objective for adaption. 
By inducing slight parameters, this linear transformation is capable of adapting our generalizable embeddings for various domains.

\section{Experiments}
In this section, we present the experimental results and corresponding analysis to answer the following research questions (\textbf{RQ}s).
\begin{itemize}[leftmargin=*]
    \item \textbf{RQ1:} How effective is our proposed \ours compared with other embedding models, including general and specialized ones?
    \item \textbf{RQ2:} How does each training stage or modification contribute to the performance of \ours?
    \item \textbf{RQ3:} What are the key properties of \ours?
\end{itemize}

\subsection{Experiment Settings}
We systematically present the details of datasets, evaluation metrics, the baselines embedding models for comparison, downstream sequential recommenders used for evaluation, as well as the implementation details.

\subsubsection{Datasets and Evaluation Metrics}
We elaborate on the selected datasets for pre-training embedding model and downstream sequential recommendation tasks, followed by descriptions of the evaluation metrics.

\textbf{Pre-training datasets.} 
Following prior works \cite{BLAIR, EasyRec}, our embedding model is pre-trained on a mixture of six datasets collected from the Amazon platform \cite{BLAIR}. These datasets span diverse categories, including \textit{Video Games} (\textbf{Games}), \textit{Arts, Crafts, and Sewing} (\textbf{Arts}), \textit{Movies and TV} (\textbf{Movies}), \textit{Home and Kitchen} (\textbf{Home}), \textit{Electronics} (\textbf{Electronics}), and \textit{Tools and Home Improvement} (\textbf{Tools}). The datasets consist of user interactions spanning from June 1996 to September 2023\footnote{\url{https://amazon-reviews-2023.github.io/}}. For all six training datasets, we apply 5-core filtering and limit the maximum historical interaction sequence length to 10. The datasets are partitioned into training, validation, and test sets using the leave-one-out strategy, where the last two interactions in each user sequence are reserved for validation and testing, respectively. Only the training data is used for pre-training \ours, while the validation and test sets are reserved for downstream evaluation. Detailed statistics of each pre-training dataset are listed in Table~\ref{tab:pretrain_dataset_statistics}.

\begin{table}[t]
\begin{center}
\caption{The statistics of the pre-training datasets.}
\label{tab:pretrain_dataset_statistics}
\vspace{-0.1in}
    \begin{tabular}{lcccccc}
        \toprule
        \textbf{Dataset} & \textbf{\#Items}  & \textbf{\#Interactions} \\
        \midrule
         Games & 9,517 & 153,221 \\
         Arts & 12,454 & 132,566 \\
         Movies & 13,190 & 136,471 \\
        Home & 33,478 & 256,001 \\
        Electronics & 20,150 & 197,984 \\
        Tools & 19,964 & 159,969 \\
        \midrule
        \textbf{Total} & 108,753 & 	1,035,212 \\
        
        \bottomrule
    \end{tabular}
\end{center}
\vspace{-0.25in}
\end{table}

\begin{table}[t]
\begin{center}
\caption{The statistics of the in-domain and out-of-domain datasets used in downstream sequential recommendation.}
\label{tab:downstream_dataset_statistics}
\vspace{-0.1in}
\resizebox{0.48\textwidth}{!}{
    \begin{tabular}{lcccccc}
        \toprule
        \textbf{Dataset} & \textbf{\#Items}  & \textbf{\#Interactions} & \textbf{Train} & \textbf{Val\&Test}\\
        \midrule
         Games & 9,517 & 153,221 & 122,577 & 15,322 \\
         Arts & 12,454 & 132,566 & 106,052 & 13,257 \\
         Movies & 13,190 & 136,471 & 109,177 & 13,647 \\
        \midrule
        Sports & 13,952 & 136,740 & 109,392 & 13,674 \\
        Baby & 6,837 & 97,899 & 78,319 & 9,790 \\
        Goodreads & 4,550 & 158,347 & 137,069 & 10,639 \\
        \bottomrule
    \end{tabular}
}
\end{center}
\vspace{-0.25in}
\end{table}

\textbf{Downstream sequential recommendation datasets.} For downstream sequential recommenders utilizing the generated embeddings, we train and test the downstream recommenders on the same datasets with identical data splits as used in pre-training. Specifically, we present experimental results on three datasets: \textbf{Games}, \textbf{Arts}, and \textbf{Movies}. To further evaluate the generalization ability of \ours to unseen domains, we additionally include three more out-of-domain datasets that differ significantly from the pre-training data. Specifically, we select \textbf{Sports}, \textbf{Baby} \cite{BLAIR} from Amazon, which contain items from categories absent in the training set. We further evaluate our embedding model on cross-platform dataset: \textbf{Goodreads}\footnote{\url{https://cseweb.ucsd.edu/~jmcauley/datasets/goodreads.html}} \cite{Goodreads_1, Goodreads_2}. Detailed statistics of downstream sequential recommendation datasets are presented in Table~\ref{tab:downstream_dataset_statistics}. 

\textbf{Evaluation Mertrics}. We follow the prior works \cite{LightGCN, BLAIR, EasyRec} and perform full ranking with all items in the dataset as potential candidates during evaluation. The performance of the recommenders is evaluated with the Recall@$k$ and NDCG@$k$, where $k \in \{10, 20\}$. 
To ensure the reliability of experimental results and mitigate the impact of unavoidable randomness during downstream recommender training, all reported performances in this section are averaged over three runs with different random seeds.

\begin{table*}[t]
\caption{Performance comparison of different embedding methods under in-domain and out-of-domain datasets. R is shorts for Recall, N is short for NDCG, and \%Improv. indicates the relative improvement compared to the strongest baselines.}
\vspace{-0.1in}
\label{tab:overall_performance}
\centering
\setlength{\tabcolsep}{1mm}{
    \resizebox{0.93\textwidth}{!} {
        \begin{tabular}{ll cccc cccc cccc}
        \toprule

        \multicolumn{14}{c}{\textbf{In-Domain Datasets}} \\
         \cmidrule(lr){1-14} \multicolumn{2}{c}{\multirow{2.4}{*}{Models}} & \multicolumn{4}{c}{Games} & \multicolumn{4}{c}{Arts} & \multicolumn{4}{c}{Movies} \\
        \cmidrule(lr){3-6} \cmidrule(lr){7-10} \cmidrule(lr){11-14} & & R@10 & N@10 & R@20 & N@20 & R@10 & N@10 & R@20 & N@20 & R@10 & N@10 & R@20 & N@20 \\
        \midrule
         
        \multirow{8}{*}{\textbf{GRU4Rec}} & BERT & 0.0365 & 0.0184 & 0.0573 & 0.0236 & 0.0363 & 0.0191 & 0.0559 & 0.0240 & 0.0243 & 0.0126 & 0.0383 & 0.0160 \\
        & GTE & 0.0540 & 0.0290 & 0.0792 & 0.0353 & 0.0348 & 0.0185 & 0.0569 & 0.0240 & 0.0396 & 0.0195 & 0.0583 & 0.0242\\
        & BGE & 0.0491 & 0.0261 & 0.0760 & 0.0329 & 0.0413 & 0.0221 & 0.0632 & 0.0276 & 0.0379 & 0.0187 & 0.0587 & 0.0239 \\
        & LLM2Vec & 0.0540 & 0.0286 & 0.0784 & 0.0348 & 0.0473 & 0.0274 & 0.0678 & 0.0325 & 0.0370 & 0.0187 & 0.0557 & 0.0234 \\
        & BLAIR & 0.0455 & 0.0245 & 0.0713 & 0.0309 & 0.0416 & 0.0233 & 0.0639 & 0.0289 & 0.0379 & 0.0188 & 0.0583 & 0.0239  \\
        & EasyRec & 0.0450 & 0.0235 & 0.0700 & 0.0298 & 0.0436 & 0.0232 & 0.0643 & 0.0284 & 0.0356 & 0.0180 & 0.0551 & 0.0229 \\
        & LLMEmb & 0.0544 & 0.0298 & 0.0775 & 0.0357 & 0.0480 & 0.0277 & 0.0673 & 0.0325 & 0.0377 & 0.0196 & 0.0538 & 0.0236 \\
        \cmidrule{2-14}
        & \textbf{\ours} & \textbf{0.0624} & \textbf{0.0344} & \textbf{0.0874} & \textbf{0.0408} & \textbf{0.0590} & \textbf{0.0366} & \textbf{0.0802} & \textbf{0.0419} & \textbf{0.0419} & \textbf{0.0214} & \textbf{0.0595} & \textbf{0.0258} \\
        & \%Improv. & 14.76\% & 15.46\% & 10.35\% & 14.31\% & 22.83\% & 32.32\% & 18.16\% & 29.03\% & 5.92\% & 9.46\% & 1.46\% & 6.77\% \\
        \midrule

        \multirow{8}{*}{\textbf{SASRec}} & BERT & 0.0585 & 0.0311 & 0.0863 & 0.0381 & 0.0650 & 0.0405 & 0.0869 & 0.0460 & 0.0447 & 0.0240 & 0.0646 & 0.0290\\
        & GTE & 0.0641 & 0.0349 & 0.0911 & 0.0418 & 0.0644 & 0.0394 & 0.0880 & 0.0454 & 0.0570 & 0.0300 & 0.0817 & 0.0363 \\
        & BGE & 0.0733 & 0.0410 & 0.1022 & 0.0483 & 0.0748 & 0.0475 & 0.1006 & 0.0540 & 0.0626 & 0.0350 & 0.0847 & 0.0406 \\
        & LLM2Vec & 0.0740 & 0.0407 & 0.1029 & 0.0480 & 0.0770 & 0.0506 & 0.1007 & 0.0566 & 0.0662 & 0.0384 & 0.0874 & 0.0438 \\
        & BLAIR & 0.0654 & 0.0361 & 0.0954 & 0.0437 & 0.0648 & 0.0379 & 0.0906 & 0.0444 & 0.0581 & 0.0315 & 0.0801 & 0.0370 \\
        & EasyRec & 0.0647 & 0.0357 & 0.0926 & 0.0428 & 0.0658 & 0.0395 & 0.0929 & 0.0463 & 0.0528 & 0.0278 & 0.0739 & 0.0331 \\
        & LLMEmb & 0.0813 & 0.0487 & 0.1085 & 0.0555 & 0.0865 & 0.0601 & 0.1086 & 0.0657 & 0.0659 & 0.0390 & 0.0837 & 0.0435 \\
        \cmidrule{2-14}
        & \textbf{\ours} & \textbf{0.0865} & \textbf{0.0521} & \textbf{0.1157} & \textbf{0.0595} & \textbf{0.0925} & \textbf{0.0637} & \textbf{0.1142} & \textbf{0.0692} & \textbf{0.0705} & \textbf{0.0429} & \textbf{0.0895} & \textbf{0.0477} \\
        & \%Improv. & 6.42\% & 6.92\% & 6.70\% & 7.04\% & 6.95\% & 5.97\% & 5.16\% & 5.29\% & 6.49\% & 10.01\% & 2.40\% & 9.13\% \\

        \cmidrule(lr){1-14} \multicolumn{14}{c}{\textbf{Out-Of-Domain Datasets}} \\

        \cmidrule(lr){1-14} \multicolumn{2}{c}{\multirow{2.4}{*}{Models}} & \multicolumn{4}{c}{Sports} & \multicolumn{4}{c}{Baby} & \multicolumn{4}{c}{Goodreads} \\
        \cmidrule(lr){3-6} \cmidrule(lr){7-10} \cmidrule(lr){11-14} & & R@10 & N@10 & R@20 & N@20 & R@10 & N@10 & R@20 & N@20 & R@10 & N@10 & R@20 & N@20 \\
        \midrule
         
        \multirow{8}{*}{\textbf{GRU4Rec}} & BERT & 0.0335 & 0.0183 & 0.0499 & 0.0224 & 0.0111 & 0.0050 & 0.0252 & 0.0086 & 0.0851 & 0.0412 & 0.1226 & 0.0506 \\
        & GTE & 0.0295 & 0.0147 & 0.0459 & 0.0188 & 0.0226 & 0.0116 & 0.0340 & 0.0145 & 0.1169 & 0.0599 & 0.1701 & 0.0733 \\
        & BGE & 0.0489 & 0.0281 & 0.0685 & 0.0330 & 0.0252 & 0.0131 & 0.0364 & 0.0158 & 0.1072 & 0.0585 & 0.1517 & 0.0697 \\
        & LLM2Vec & 0.0663 & 0.0464 & 0.0810 & 0.0501 & 0.0254 & 0.0138 & 0.0362 & 0.0165 & 0.1174 & 0.0655 & 0.1643 & 0.0773 \\
        & BLAIR & 0.0537 & 0.0316 & 0.0735 & 0.0366 & 0.0207 & 0.0099 & 0.0316 & 0.0127 & 0.0939 & 0.0496 & 0.1339 & 0.0596 \\
        & EasyRec & 0.0492 & 0.0270 & 0.0674 & 0.0315 & 0.0207 & 0.0105 & 0.0275 & 0.0122 & 0.0951 & 0.0477 & 0.1364 & 0.0581 \\
        & LLMEmb & 0.0705 & 0.0482 & 0.0861 & 0.0521 & 0.0252 & 0.0136 & 0.0378 & 0.0168 & 0.1219 & 0.0701 & 0.1667 & 0.0814 \\
        \cmidrule{2-14}
        & \textbf{\ours} & \textbf{0.0828} & \textbf{0.0632} & \textbf{0.0948} & \textbf{0.0662} & \textbf{0.0327} & \textbf{0.0181} & \textbf{0.0463} & \textbf{0.0216} & \textbf{0.1299} & \textbf{0.0761} & \textbf{0.1738} & \textbf{0.0872} \\
        & \%Improv. & 17.50\% & 31.18\% & 10.07\% & 27.06\% & 28.55\% & 31.61\% & 22.32\% & 28.51\% & 6.58\% & 8.68\% & 2.17\% & 7.15\% \\
        \midrule

        \multirow{8}{*}{\textbf{SASRec}} & BERT & 0.0860 & 0.0649 & 0.1017 & 0.0689 & 0.0114 & 0.0050 & 0.0232 & 0.0080 & 0.1479 & 0.0858 & 0.1929 & 0.0972 \\
        & GTE & 0.0823 & 0.0584 & 0.1001 & 0.0629 & 0.0264 & 0.0142 & 0.0387 & 0.0173 & 0.1488 & 0.0851 & 0.1944 & 0.0967 \\
        & BGE & 0.0974 & 0.0736 & 0.1141 & 0.0778 & 0.0428 & 0.0250 & 0.0569 & 0.0286 & 0.1445 & 0.0813 & 0.1972 & 0.0945 \\
        & LLM2Vec & 0.1079 & 0.0854 & 0.1234 & 0.0893 & 0.0561 & 0.0339 & 0.0722 & 0.0379 & 0.1424 & 0.0790 & 0.1906 & 0.0911 \\
        & BLAIR & 0.0893 & 0.0614 & 0.1091 & 0.0664 & 0.0332 & 0.0180 & 0.0484 & 0.0218 & 0.1508 & 0.0860 & 0.2000 & 0.0984 \\
        & EasyRec & 0.0887 & 0.0627 & 0.1061 & 0.0671 & 0.0271 & 0.0154 & 0.0381 & 0.0182 & 0.1445 & 0.0825 & 0.1908 & 0.0941 \\
        & LLMEmb & 0.1131 & 0.0936 & 0.1257 & 0.0969 & 0.0659 & 0.0439 & 0.0807 & 0.0476 & 0.1374 & 0.0778 & 0.1838 & 0.0895 \\
        \cmidrule{2-14}
        & \textbf{\ours} & \textbf{0.1170} & \textbf{0.0976} & \textbf{0.1289} & \textbf{0.1006} & \textbf{0.0708} & \textbf{0.0503} & \textbf{0.0850} & \textbf{0.0539} & \textbf{0.1530} & \textbf{0.0897} & \textbf{0.2017} & \textbf{0.1020} \\
        & \%Improv. & 3.51\% & 4.26\% & 2.56\% & 3.89\% & 7.39\% & 14.56\% & 5.32\% & 13.04\% & 1.45\% & 4.37\% & 0.83\% & 3.73\% \\        

        \bottomrule
        \end{tabular}
    }
}
\vspace{-0.20in}
\end{table*}

\subsubsection{Baselines and Downstream Recommenders} 
For baseline embedding models, we compare \ours with a diverse set of models, including both general-purpose and recommendation-specific models. The general embedding models include \textbf{BERT}, \textbf{GTE} \cite{GTE}, \textbf{BGE} \cite{BGE-li2024}, and \textbf{LLM2Vec} \cite{LLM2Vec}. The recommendation-specific models include \textbf{EasyRec} \cite{EasyRec}, \textbf{BLAIR} \cite{BLAIR}, and \textbf{LLMEmb} \cite{LLMEmb}.

These embedding models provide better initialization for item representation learning and can be integrated into various sequential recommenders. To evaluate their effectiveness, we test these embedding models on two different sequential recommender architectures: \textbf{GRU4Rec} \cite{GRU4Rec}, and \textbf{SASRec} \cite{SASRec}. 
We leave detailed introductions to each embedding model and downstream sequential recommenders in Appendix~\ref{subsec:baseline_embedding_models}.

\subsubsection{Implementation Details}
Our \ours is initialized with a pre-trained LLM. Unless otherwise specified, all experimental results reported for \ours in this paper are based on the Qwen2-0.5B backbone.
In the first collaborative supervised fine-tuning stage, we utilize the AdamW optimizer \cite{AdamW} with learning rate set to \(3e\!-\!4\). The model is fully fine-tuned with all parameters open for 10,000 steps with the effective batch size set to 128. Then for the masked next token prediction, following the established setting of prior works \cite{BERT, LLM2Vec}, we randomly mask 20\% of the input tokens and adopt the same hyperparameter settings as LLM2Vec \cite{LLM2Vec}. The model is only fine-tuned for 1,000 steps with the effective batch size set to 32, which takes less than 2 hours on one single Nvidia A40 GPU. Finally, for item-level contrastive learning, item representations are augmented with dropout rate set to 0.2 and contrastive learning temperature $\tau$ set to 0.2. The model is optimized with AdamW optimizer for 1,000 steps with learning rate set to \(2e\!-\!4\) and effective batch size set to 256.

For downstream recommenders, all models are trained using cross-entropy loss with AdamW optimizer. To ensure fair comparisons across different embedding models, we use a fixed set of hyperparameters for each recommender while varying only the text embeddings. The learning rate is set to \(1e\!-\!3\) for SASRec and \(1e\!-\!4\) for GRU4Rec. Across all three recommenders, the weight decay is fixed at \(1e\!-\!4\), dropout out rate at $0.3$, and the extracted text embeddings are projected to 128 dimensions using a trainable linear layer. The recommenders are trained for up to 500 epochs with an early stopping mechanism, which terminates training if the validation performance does not improve for 20 consecutive epochs.
Experiments in this paper are conducted on 4 Nvidia A40 (48G) GPUs.

\subsection{Performance Comparison (\textbf{RQ1})}
We evaluate the effectiveness of our proposed \ours using two downstream sequential recommenders on three in-domain and three out-of-domain datasets. The comprehensive results, presented in Table~\ref{tab:overall_performance}, reveal the following key observations.

\ours consistently outperforms all baseline models across both recommenders on all datasets. For the in-domain datasets, it achieves an average relative improvement of 15\% on Games and Arts, and over 7\% on Movies. The significant performance gain has well demonstrated the \ours's ability to effectively capture collaborative filtering signals that significantly enhance sequential recommendation performance. More importantly, \ours also excels on out-of-domain datasets, which consist of item categories absent from the pre-training data. Even on Goodreads, which differs from the training set in both item categories and source platform, \ours maintains a moderate yet consistent performance gain. The strong results on out-of-domain datasets indicate that training on a diverse set of recommendation datasets can bring both CF awareness and generalization ability to unseen domains. These findings highlight the potential of LLMs as generalizable embedding models for recommendation.

For general text embedding models, the latest methods surpass earlier models like BERT due to their enhanced semantic understanding capabilities. In contrast, recommendation-specific embedding models such as BLAIR and EasyRec benefit from learning CF signals, allowing them to outperform general text embeddings like BERT. However, their language backbones with limited semantic understanding constrain their effectiveness, resulting in lower performance compared to more advanced models like BGE and LLM2Vec. 
LLMEmb inherits the strong semantic understanding of LLMs and further achieves performance gains by fine-tuning on recommendation-specific tasks.
These results demonstrate that both comprehensive semantic comprehension and CF awareness are crucial for improving downstream recommenders, underscoring the importance of developing a powerful embedding model integrating both aspects.

\subsection{Ablation Study (\textbf{RQ2})} \label{subsec:ablation_study}

\begin{table}[t]
\begin{center}
\caption{The ablation study of \ours.}
\label{tab:ablation_study}
\vspace{-0.1in}
\resizebox{0.49\textwidth}{!}{
    \begin{tabular}{lcccccc}
        \toprule
         \multirow{2.4}{*}{Models} & \multicolumn{2}{c}{Games} & \multicolumn{2}{c}{Sports} & \multicolumn{2}{c}{Goodreads} \\
         \cmidrule(lr){2-3} \cmidrule(lr){4-5} \cmidrule(lr){6-7} & R@10 & N@10 & R@10 & N@10 & R@10 & N@10 \\
        \midrule
         Casual & 0.0373 & 0.0201 & 0.0167 & 0.0091 & 0.0724 & 0.0375 \\
         Bidirectional & 0.0740 & 0.0407 & 0.1079 & 0.0854 & 0.1424 & 0.0790 \\
        \midrule
         CSFT & 0.0795 & 0.0472 & 0.1119 & 0.0935 & 0.1513 & 0.0882 \\
         IEM$_{1}$ (MNTP) & 0.0801 & 0.0477 & 0.1147 & 0.0956 & 0.1564 & 0.0916 \\
         IEM$_{2}$ (IC) & 0.0865 & 0.0521 & 0.1170 & 0.0976 & 0.1530 & 0.0897 \\
        \bottomrule
    \end{tabular}
}
\end{center}
\vspace{-0.25in}
\end{table}

\hyz{Do there need to be model between causal and bidrectional?}
\hyz{Do I need to point out that results are from SASRec in table title?}

To analyze the contribution of each training stage to the performance of \ours, we conduct an ablation study using the strongest sequential recommender, SASRec, as the fixed downstream model. The evaluation covers one in-domain dataset (Games) and two out-of-domain datasets (Sports and Goodreads). The detailed results are presented in Table~\ref{tab:ablation_study}.
``Causal'' and ``Bidirectional'' present the performance of our backbone LLM, Qwen2-0.5B, under different embedding generation strategies. In the causal setting, item embeddings are derived from the last hidden state of the [EOS] token while maintaining the causal attention mask. In the bidirectional setting, embeddings are obtained by average pooling the last hidden states of all tokens in the item title. Experimental results demonstrate that bidirectional attention consistently outperforms causal attention. While causal attention is beneficial for language generation tasks by conditioning later tokens on prior ones, it proves suboptimal for embedding generation, as it limits the model’s ability to capture comprehensive contextual representations.

The bottom half of Table~\ref{tab:ablation_study} highlights the impact of each training stage on the performance of \ours. ``CSFT'' represents the model's performance after Collaborative Supervised Fine-Tuning (CSFT). Item-level Embedding Modeling (IEM) consists of two steps: ``IEM$_{1}$ (MNTP)'' indicates the performance after additional Masked Next-Token Prediction (MNTP) training, and IEM$_{2}$ (IC) reflects the performance after item-level contrastive learning (IC), which ultimately results in \ours.
Experimental results indicate that collaborative supervised fine-tuning (CSFT) contributes the most significant performance improvement across both in-domain and out-of-domain datasets, highlighting the importance of capturing CF signals in sequential recommendation and also demonstrating the effectiveness of CSFT. In comparison, masked next-token prediction (MNTP) provides a smaller yet consistent performance boost, suggesting its role in improving bidirectional contextual representations. Finally, item-level contrastive learning further enhances performance, yielding substantial gains on in-domain datasets and moderate improvements on out-of-domain datasets.

\subsection{Model Study (\textbf{RQ3})}

\begin{figure*}[t]
    \centering
    \includegraphics[width = 0.88\linewidth]{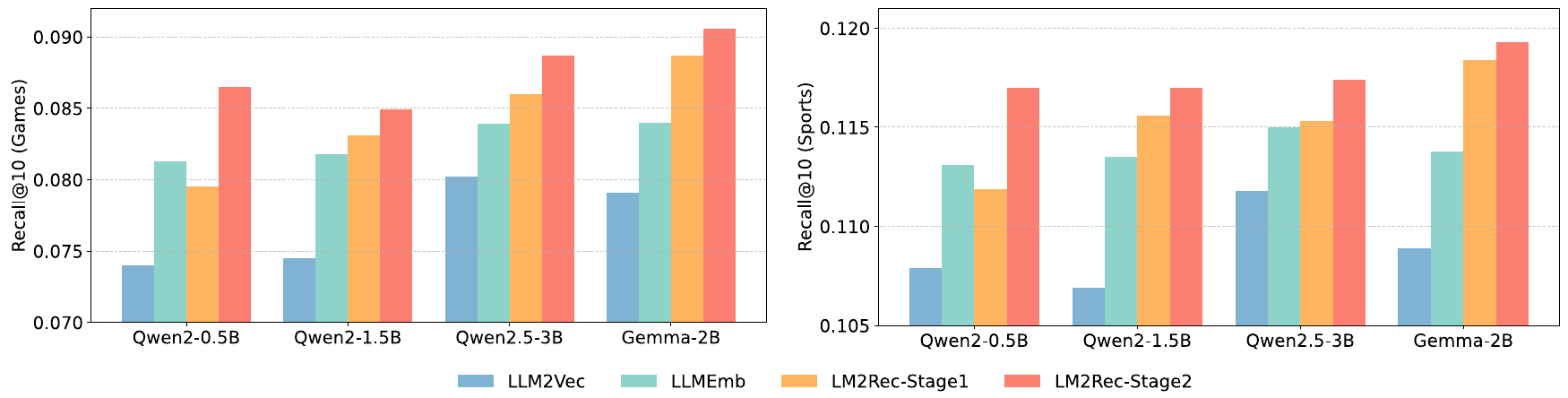}
    \vspace{-0.12in}
    \caption{Performance comparison of embedding methods across different LLM backbones.}
    \label{fig:model_study_llm_backbones}
    \vspace{-0.18in}
\end{figure*}

\subsubsection{Effect of Different LLM Backbones}
\modified{
As \ours builds upon pre-trained LLMs, the choice of backbone naturally affects its performance. To examine this effect, we evaluate LLM2Rec on various LLM backbones. Specifically, for each LLM backbone, we evaluate the quality of the generated embedding after two training stages of \ours, \ie the collaborative supervised fine-tuning (LLM2Rec-Stage1) and item-level embedding modeling (LLM2Rec-Stage2). 
To provide a comprehensive comparison, we further include two baseline methods: LLM2Vec, a state-of-the-art general-purpose embedding model, and LLMEmb, a sequential recommendation-specific embedding model. Since \ours, LLM2Vec, and LLMEmb are all fine-tuned from a pre-trained LLM, we compare their performance under the same backbone LLM.
All embeddings are integrated into the same downstream recommender, SASRec, and tested on both the in-domain dataset (Games) and the out-of-domain dataset (Sports).

As the results shown in Figure~\ref{fig:model_study_llm_backbones}, models built upon stronger LLM backbones generally achieve better performance on both the in-domain (Games) and out-of-domain (Sports) datasets. This trend is consistent with expectations, as larger and more powerful LLMs tend to possess stronger semantic understanding and greater generalization capabilities. Across all evaluated backbones, both training stages of LLM2Rec contribute to consistent improvements in the effectiveness of the generated embeddings for recommendation tasks. Notably, \ours consistently outperforms the general-purpose embedding baseline, \ours, on a variety of LLM backbones. Furthermore, after completing both stages of training, \ours (LLM2Rec-Stage2) also surpasses the recommendation-specific embedding method, LLMEmb. The results on four different LLMs confirm the effectiveness of \ours and its robustness in generalizing across different LLM backbones. It is also notable that \ours achieves competitive performance when built on the lightweight Qwen2-0.5B model, offering a favorable trade-off between effectiveness and computational cost compared to larger backbones.
}

\subsubsection{Effect of Mixed Dataset Training}

\begin{table}[t]
\begin{center}
\caption{Effect of mixed training dataset. ID is short for in-domain and OOD is short for out-of-domain.}
\label{tab:model_study_mixed_datasets}
\vspace{-0.1in}
\resizebox{0.49\textwidth}{!}{
    \begin{tabular}{lcccccc}
        \toprule
         
         \multirow{2.4}{*}{Models} & \multicolumn{2}{c}{Games ({ID})} & \multicolumn{2}{c}{Sports ({OOD})} & \multicolumn{2}{c}{Goodreads ({OOD})} \\
         \cmidrule(lr){2-3} \cmidrule(lr){4-5} \cmidrule(lr){6-7} & R@10 & N@10 & R@10 & N@10 & R@10 & N@10 \\
        \midrule
         Backbone & 0.0740 & 0.0407 & 0.1079 & 0.0854 & 0.1424 & 0.0790 \\
        \midrule
         Single & 0.0857 & 0.0500 & 0.1099 & 0.0921 & 0.1473 & 0.0833 \\
         \%Improv. & 15.81\% & 22.90\% & 1.83\% & 7.76\% & 3.41\% & 5.44\% \\
        \midrule
         Mix-2 & 0.0856 & 0.0517 & 0.1111 & 0.0936 & 0.1493 & 0.0862 \\
         \%Improv. & 15.66\% & 26.84\% & 2.96\% & 9.52\% & 4.82\% & 9.12\%  \\
         \midrule
         Mix-6 & 0.0795 & 0.0472 &  0.1119 & 0.0935 & 0.1513 & 0.0882\\
         \%Improv. & 7.38\% & 15.80\% & 3.68\% & 9.45\% & 6.20\% & 11.66\% \\
        \bottomrule
    \end{tabular}
}
\end{center}
\vspace{-0.2in}
\end{table}

Our proposed \ours is pre-trained on a mixture of six datasets spanning different categories. To examine the impact of mixed pre-training datasets, we evaluate embedding models trained on a single dataset, a mixture of two datasets, and a mixture of six datasets. For simplicity, we omit the item-level embedding modeling steps (masked next-token prediction and item-level contrastive learning), and all results are reported using SASRec as the fixed downstream recommender. 
We define four settings: ``Backbone'', which represents the base performance of the backbone LLM, Qwen2-0.5B; ``Single'', where the model is pre-trained only on the Games dataset; ``Mix-2'', pre-trained on a combination of Games and Arts; and ``Mix-6'', which corresponds to \ours, pre-trained on all six datasets. Among the three evaluated datasets, Games is in-domain for all models, whereas Sports and Goodreads are out-of-domain, as they are excluded from the pre-training datasets.  

Experimental results as shown in Table~\ref{tab:model_study_mixed_datasets} indicate that pre-training on a diverse set of datasets improves the model’s ability to generalize to unseen domains, leading to more robust embeddings across different datasets. Meanwhile, for in-domain data, pre-training on a concentrated subset of item categories yields higher accuracy, suggesting that category-specific pre-training can be beneficial for domain-specific recommendations.

\hyz{For simplicity, we ignore the item-level embedding modeling step...  Is it risky?}

\subsubsection{Efficiency Analysis}

\begin{figure}[t]
    \centering
    \includegraphics[width = 0.98\linewidth]{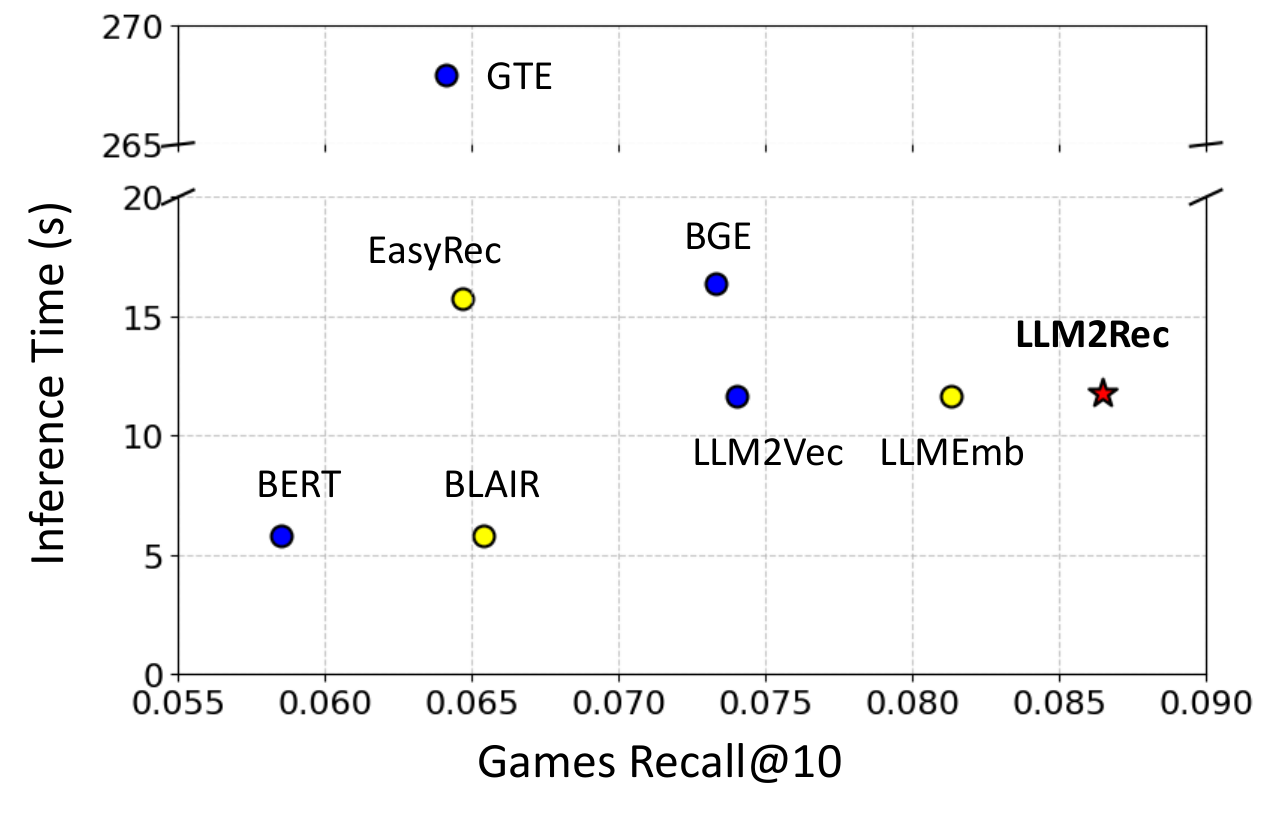}
    \vspace{-0.2in}
    \caption{Comparison of inference time and performance of different embedding models.}
    \label{fig:efficiency}
    \vspace{-0.35in}
\end{figure}

In addition to effectiveness, the efficiency of embedding models is crucial for practical deployment in sequential recommender systems. We measure the inference time required by each embedding model to encode all item titles in the Games dataset (comprising 9,517 items) on a single Nvidia A40 GPU. The results are shown in Figure~\ref{fig:efficiency}. The vertical axis represents the total inference time, while the horizontal axis shows the Recall@10 performance of SASRec under different embedding models.

The smallest embedding models, BERT and BLAIR, take the shortest inference times, with BLAIR achieving a notable performance improvement over BERT due to its specialized fine-tuning for recommendation tasks. The latest embedding models generally deliver better performance, driven by enhanced semantic understanding, but at the cost of higher computational overhead. Larger models, such as GTE, which leverages a 7B-parameter backbone, suffer from significantly increased inference time. LLM2Vec and LLMEmb reported in Figure~\ref{fig:efficiency} are built on the lightweight yet powerful Qwen2-0.5B model. They achieve strong performance with relatively low computational cost, demonstrating a favorable balance between effectiveness and efficiency. Our proposed \ours inherits the efficiency of Qwen2-0.5B while enhancing performance through recommendation-specific fine-tuning. Overall, \ours generates more effective embeddings for downstream sequential recommenders while maintaining practical inference efficiency.

\section{Conclusion \& Discussion}

In this work, we introduced \ours, a specialized embedding model for sequential recommendation that incorporates both the comprehensive semantic understanding of LLMs and awareness of CF signals. General embedding models fail to capture the latent item relationships, \ie CF signals, resulting in suboptimal performance in sequential recommendation tasks.  
\ours bridges this gap by leveraging a two-stage training framework including collaborative supervised fine-tuning (CSFT) and item-level embedding modeling (IEM). CSFT fine-tunes the LLM to capture the CF signals with user interaction sequences and IEM further transforms the decoder-only LLM into embedding model focusing on item embedding generation for sequential recommendation.
Extensive experiments on real-world datasets demonstrate that \ours consistently outperforms strong baseline embedding models across both in-domain and out-of-domain recommendation tasks. Notably, it achieves significant improvements while maintaining computational efficiency. Our results highlight the potential of LLMs as powerful embedding models for sequential recommendation.

While \ours demonstrates strong effectiveness, several promising research directions remain.  
First, real-world user interaction data from e-commerce platforms often contain substantial noise. Enhancing robustness through noise filtering or data augmentation could further improve \ours.  
Second, due to computational constraints, this study evaluates LLM2Rec on LLM backbones with up to 3B parameters. Despite its effectiveness, exploring larger-scale LLMs under the \ours framework could unlock further performance gains.  
Third, experimental results indicate that training on mixed datasets enhances generalization. Constructing a more diverse dataset with items from multiple platforms and categories could improve generalization, advancing toward a universal recommender system trained once and deployed widely.
\section*{acknowledgement}
This research is supported by NExT++ Research Center.

\bibliographystyle{ACM-Reference-Format}
\balance
\bibliography{references}

\clearpage
\appendix

\section{Implemental Details}

\subsection{Training Pseudo-code.}
Here we provide the training pseudo-code, indicating that LLM2Rec is sequentially optimized with different stages, as shown in Algorithm~\ref{alg:training}. 
The training starts from collaborative supervised fine-tuning to item-level embedding modeling. Due to the different objectives, we employ different sampling at different stages. Specifically, the user interaction sequence and the next item pairs constitute the samples for the collaborative supervised fine-tuning, while single items serve for the item-level embedding modeling. Additionally, during the item-level contrastive learning, we augment the items to two views, making it an extra sampling that differs from the training with MNTP objective.


\begin{algorithm}
\caption{Training Strategy for LLM2Rec}
\label{alg:training}
\begin{algorithmic}[1]
    \Require Training dataset $\mathcal{D}^{\text{train}}$, pretrained LLM $\mathcal{E}$ with parameter $\theta$, learning rate $\eta$, epochs $E$, temperature $\tau$
    
    \State \textbf{Initialization:} Load pretrained LLM $\mathcal{E}$, Initialize optimizer
    
    \State \textbf{Stage 1: Collaborative Supervised Fine-tuning}
    \For{epoch $= 1$ to $E_1$}
        \For{each batch  $\{(X_u, i_{N_u + 1}\}) \subset \mathcal{D}^{\text{train}}$}
            \State Encode item sequence: $\mathbf{t}_u \gets \text{Tokenize}(X_u)$
            \State Compute loss $\mathcal{L}_{\text{CSFT}}$ in Equation~\ref{eq:csft}
            \State Update model parameters: $\theta_{\text{1}} \gets \theta - \eta \nabla_{\theta} \mathcal{L}_{\text{CSFT}}$
        \EndFor
    \EndFor

    \State \textbf{Stage 2: Item-level Embedding Modeling}
    \State \textbf{Step 2.1: Reform LLM with Bidirectional Attention}
    \For{epoch $= 1$ to $E_2$}
        \For{each batch $\{i\} \subset \mathcal{I}$}
            \State Tokenize item description: $\mathbf{t}_i$
            \State Randomly mask tokens for MNTP
            \State Compute loss $\mathcal{L}_{\text{MNTP}}$ in Equation~\ref{eq:mntp}
            \State Update model parameters: $\theta' \gets \theta - \eta \nabla_{\theta_{\text{1}}} \mathcal{L}_{\text{MNTP}}$
        \EndFor
    \EndFor

    \State \textbf{Step 2.2: Item-level Contrastive Learning}
    \For{epoch $= 1$ to $E_3$}
        \For{each batch $\{i\} \subset \mathcal{I}$}
            \State Generate two masked views: $\tilde{\mathbf{t}}_i^1, \tilde{\mathbf{t}}_i^2$
            \State Compute item embeddings: $\mathbf{z}_i^1 = \mathcal{E}(\tilde{\mathbf{t}}_i^1)$, $\mathbf{z}_i^2 = \mathcal{E}(\tilde{\mathbf{t}}_i^2)$
            \State Compute loss $\mathcal{L}_{\text{IC}}$ in Equation~\ref{eq:ic}
            \State Update model parameters: $\tilde{\theta} \gets \theta' - \eta \nabla_{\theta'} \mathcal{L}_{\text{IC}}$
        \EndFor
    \EndFor

    \State \Return LLM2Rec $\mathcal{E}$ with parameter $\tilde{\theta}$.

\end{algorithmic}
\end{algorithm}


        

\subsection{Baselines and Sequential Recommenders.}
\label{subsec:baseline_embedding_models}
This section provides a brief introduction to each baseline embedding model used in our experiments.

\begin{itemize}[leftmargin=*]
\item \textbf{BERT}~\cite{BERT} is a milestone embedding model pre-trained using mask language modeling and next sentence prediction. Built upon the transformer architecture with bidirectional attention, BERT effectively captures contextual dependencies in text, enabling more accurate semantic representations.

\item \textbf{BGE}~\cite{BGE-li2024} is a state-of-the-art embedding model built on a bidirectional Transformer architecture. Pre-trained on diverse datasets, it excels in retrieval and reranking tasks. In this paper, we use the pre-trained model BAAI/bge-large-en-v1.5 from the Hugging Face repository.

\item \textbf{GTE}~\cite{GTE} transforms the LLM into an embedding model with multi-stage contrastive learning. Specifically, we select the pre-trained model Alibaba-NLP/gte-Qwen2-7B-instruct which exhibits the highest performance on Massive Text Embedding Benchmark (MTEB) \cite{MTEB}. 
 
\item \textbf{LLM2Vec}~\cite{LLM2Vec} aims to effectively adapt pre-trained LLMs into embedding models through sentence-level adaptations. In this paper, we use Qwen2-0.5B as the backbone for LLM2Vec, ensuring consistency with our proposed \ours.

\item \textbf{EasyRec} \cite{EasyRec} is a recommendation-specific embedding model built on RoBERTa \cite{Roberta}. It leverages contrastive learning to capture collaborative filtering (CF) signals by aligning representations of user and item profiles. We load the pre-trained embedding from Hugging Face repo hkuds/easyrec-roberta-large.

\item \textbf{BLAIR} \cite{BLAIR} is a recommendation-specific embedding model, similar to EasyRec. It is pre-trained on a diverse collection of 33 Amazon datasets, comprising $3.08 \times 10^7$ data instances. In this paper, we load the pre-trained model from Hugging Face repo hyp1231/blair-roberta-base.

\item \textbf{LLMEmb} \cite{LLMEmb} adopts attribute-level augmentation and uses contrastive learning as a primary training objective to improve semantic understanding of LLM embeddings, particularly for representing the long-tail items. In this paper, we implement LLMEmb using the same LLM backbone, Qwen2-0.5B, to ensure consistency with \ours.
\end{itemize}

Two downstream sequential recommenders are used as evaluation to the embedding models:
\begin{itemize}[leftmargin=*]
\item \textbf{GRU4Rec} \cite{GRU4Rec} utilizes GRU modules to capture sequential dependencies within user interaction sequences. It is trained to predict the next item in a user’s sequence based on previously purchased items. We use the cross-entropy loss as optimization objective during the training process.

\item \textbf{SASRec} \cite{SASRec} is a transformer-based recommender widely used in sequential recommendation. It leverages self-attention to capture long-range dependencies in user interaction sequences, enhancing the accuracy of future interaction predictions. Cross-entropy loss is used as the optimization objective.
\end{itemize}



\end{document}